\documentclass[12pt]{article}
\usepackage{epsfig}
\usepackage{amsmath}
\usepackage{amssymb}
\usepackage{amsfonts}
\usepackage{latexsym}
\usepackage{wasysym}
\usepackage{multirow}
\usepackage{fixmath}
\usepackage{color}
\usepackage{stackrel}
\usepackage{txfonts}                                                            
\usepackage{bbm}
\usepackage{slashbox}

\def\eq{\begin{eqnarray}}
\def\en{\end{eqnarray}}
\definecolor{ForestGreen}{rgb}{0.133305, 0.545106, 0.133305}

\begin{document}

\title{
\vspace{-2.5cm}
\flushleft{\normalsize DESY 13-127} \\
\vspace{-0.35cm}
{\normalsize Edinburgh 2013/18} \\
\vspace{-0.35cm}
{\normalsize Liverpool LTH 983} \\
\vspace{-0.35cm}
{\normalsize September 2013} \\
\vspace{0.5cm}
\centering{\Large \bf The SU(3) Beta Function from Numerical Stochastic
Perturbation Theory}\\}
\author{R. Horsley$^a$, H.~Perlt$^{b,c}$, 
 P.E.L.~Rakow$^{d}$, G.~Schierholz$^{e}$ and A. Schiller$^{b}$\\[1em] 
$^a$ School of Physics and Astronomy,\\ University of Edinburgh,\\ Edinburgh
EH9 3JZ, United Kingdom\\ \\
$^b$ Institut f\"ur Theoretische Physik,\\ Universit\"at Leipzig,\\ 04103
Leipzig, Germany\\ \\ 
$^c$ Helmholtz Institut f\"ur Strahlen- und Kernphysik,\\ Universit\"at
Bonn,\\ 53115 Bonn, Germany\\ \\
$^d$ Theoretical Physics Division,\\ Department of Mathematical Sciences,\\
University of Liverpool,\\ Liverpool L69 3BX, United Kingdom\\ \\
$^e$ Deutsches Elektronen-Synchrotron DESY,\\ 22603 Hamburg, Germany}

\date{ }

\maketitle

\begin{abstract}
\noindent
The SU(3) beta function is computed from Wilson loops to
$20th$ order numerical stochastic perturbation theory. An attempt
is made to include massless fermions, whose contribution is known
analytically to $4th$ order. The question whether the theory  
admits an infrared stable fixed point is addressed.
\end{abstract}

The evolution of the running coupling $g^2(\mu)$
of nonabelian gauge theories as a function of the Euclidean momentum
scale $\mu$ is of fundamental interest. It is encoded in the
Callan-Symanzik $\beta$ function. Of particular interest is the
evolution of $g^2(\mu)$ at small momenta, which is determined by
the behavior of the $\beta$ function at large $g^2$. Various
possibilities come to mind. In the pure gauge theory the 
most plausible, and internally consistent, scenario is that $\mu$
cannot be taken lower than a certain value, $\mu_0 \leq \mu$, where 
$\mu_0$ is the `mass gap' of the theory.\footnote{This might be
the dynamically generated mass of the gluon.} In the theory with
dynamical massless fermions there is no mass gap, and nothing stops
$\mu$ from being taken to zero. Whether the $\beta$ function exibits
an infrared fixed point and $g^2(\mu)$ freezes at small scales $\mu$,
giving rise to a conformal window, is an open question though. The
third scenario is that the $\beta$ function has a pole at some finite
value of $g^2$, like that of the supersymmetric Yang-Mills
theory~\cite{Kogan:1995hn}. It will divide the theory into two 
phases, one being asymptotically free and another being strongly
coupled in the infrared, with $g^2(\mu)$ flowing to a point
$g^{*\,2}$, both from the small and large $g^2$ domain. 
In this work we shall seek an `all-order' perturbative solution to the
SU(3) $\beta$ function.

We start from rectangular $L \times T$ Wilson loops
$W(L,T)$ and the corresponding Creutz ratios $R(L,T)$,
\begin{equation}
R(L,T) = \frac{W(L,T)\, W(L-1,T-1)}{W(L,T-1)\, W(L-1,T)} \,.
\end{equation}
The lattice constant is taken to be $a=1$, if not stated otherwise. For
$T \gg L$ the Wilson loops can be written 
\begin{equation}
W(L,T) = C\, \exp{\{-E(L)\,T\}} \, , \quad E(L) = - \tilde{C}_F\,
\frac{g_V^2(L)}{L} + \left(\sigma -\frac{\pi}{12 L^2}\right)\,L \,, 
\label{EL}
\end{equation}
where $\tilde{C}_F=C_F/4\pi=1/3\pi$. The string tension $\sigma$, including
the contribution $-\pi/12 L^2$ from fluctuations of the bosonic
string~\cite{Luscher:1980ac}, is of nonperturbative origin and as such
not accessible perturbatively. We will comment on potential
nonperturbative contributions to the $\beta$ function later. The
factor $C$, with $\ln\, C \propto (L+T)$, 
drops out in the Creutz ratio. This leaves us with
\begin{equation}
\ln \, R(L,T) = \tilde{C}_F \,\left[ \frac{g_V^2(L)}{L} -
  \frac{g_V^2(L-1)}{L-1}\right] \,.
\end{equation}
If we now expand $g_V^2(L)$ and $g_V^2(L-1)$ around $\bar{L} =
\sqrt{L(L-1)}$, $g_V^2(L) = g_V^2(\bar{L}) +
g_V^{2\;\prime}(\bar{L})\,(L - \bar{L}) + \cdots$, we find 
\begin{equation}
\ln \, R(L,T) =  \tilde{C}_F \, \left[ - \frac{g_V^2(\bar{L})}{\bar{L}^2} +
  \frac{g_V^{2\;\prime}(\bar{L})}{\bar{L}}\right]  
              \equiv - \tilde{C}_F\, \frac{g_{qq}^2(\bar{L})}{\bar{L}^2} =
              - F(\bar{L}) \,,
\label{rc}
\end{equation}
up to a systematic error  
$\simeq - \tilde{C}_F\,g_V^{2\;\prime\prime\prime}(\bar{L})/24 \bar{L}$ arising
from the truncation of the Taylor series, where $F(L)$
is the force and $g_{qq}^2(L)$ the coupling 
in the $qq$ or force scheme.
The corresponding $\beta$ function, $\beta_{qq}(g_{qq}(L))$, is given by 
\begin{equation}
\frac{1}{2\,g_{qq}(L)}\,\frac{\partial\, g_{qq}^2(L)}{\partial\, \ln\,L} =
- \beta_{qq}(g_{qq}(L)) \,, 
\end{equation}
from which the running coupling $g_{qq}^2(\mu)$, with $\mu=1/L$, may be
obtained by solving  
\begin{equation}
\frac{\mu}{\Lambda_{qq}} =
\left(\beta_0\,g_{qq}^2(\mu)\right)^{\frac{\beta_1}{2\beta_0^2}}
\exp\left\{{\frac{1}{2\beta_0 g_{qq}^2(\mu)}} +
\int_0^{g_{qq}(\mu)} \mathrm{d}g \left(\frac{1}{\beta_{qq}(g)} +
\frac{1}{\beta_0\, g^3} - \frac{\beta_1}{\beta_0^2\, g}\right)\right\} \,.
\label{mu}
\end{equation}
Perturbatively, the $\beta$ function
\begin{equation}
\beta_{qq}(g) = - g^3 \left(\beta_0 + \beta_1\,g^2 + \beta_2^{qq}\,g^4 +
\cdots\right)
\end{equation}
is known to four loops~\cite{Donnellan:2010mx}. The first two
coefficients are universal, $\displaystyle \beta_0=11/(4\pi)^2$,
$\displaystyle\beta_1=102/(4\pi)^4$, while the remainder are scheme
dependent. 
 
\begin{figure}[b!]
\begin{center}
\epsfig{file=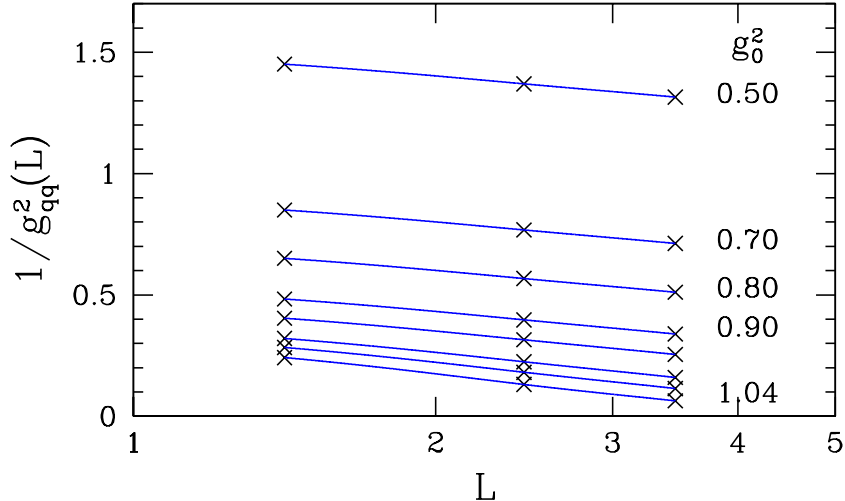,width=12.0cm,clip=}
\end{center}
\vspace*{-1cm}
\caption{A plot of the coupling $g_{qq}(L)$ as a function
  of $\ln L$. The crosses ($\times$) show the lattice results for bare
  couplings $g_0^2=0.5,\cdots, 0.9, 0.95, 1.0, 1.02, 1.04$, from top
  to bottom. The curves show a second-order Lagrange interpolation of
  the lattice data.}  
\label{fig:g}
\end{figure}

In~\cite{Horsley:2012ra} we have computed rectangular Wilson loops
$W(L,T)$ on the $12^4$ lattice for $L,T = 1,
\cdots, 6$ to $N=20$ loops in the bare coupling $g_0^2$, using numerical
stochastic perturbation theory~\cite{DiRenzo:1994sy} and the Wilson
gauge action. We did not find any evidence for factorial asymptotic
growth characteristic of an asymptotic  series and renormalon
singularities. The  
perturbative series of the smaller Wilson loops were estimated to
converge for $g_0^2 \leq 1.04$.   
Knowing the Wilson loops, we can compute the Creutz ratios. The latter
can be written
\begin{equation}
R(L,T)= 1+\sum_{n=1}^{N} r_n(L,T)\, g_0^{2n} \,,
\end{equation}
from which we obtain the running coupling
\begin{equation}
g_{qq}^2(\bar{L}) = \frac{1}{r_1(L,T)} \, \ln\, R(L,T) = g_0^2 +
\sum_{n=2}^{N} c_n(L,T)\,g_0^{2n} \,.
\label{rc2}
\end{equation}
We consider Wilson loops of size $T=5$ and $L=2, 3$ and $4$. This
leaves us with the Creutz ratios $R(2,5), R(3,5)$ and $R(4,5)$, from
which we obtain $g_{qq}^2(\bar{L})$ at $\bar{L}=\sqrt{2}, \sqrt{6}$
and $\sqrt{12}$. In first (one-loop) approximation 
$g_{qq}^{-2}(L)$ is a linear function of $\ln\, L$. In 
Fig.~\ref{fig:g} we plot $g_{qq}^{-2}(L)$ against $\ln L$ for
various values of $g_0^2$. At $g_0^2=1.04$ we find
$g_{qq}^2(\sqrt{12}) \approx 16$, which allows us to probe rather
large values of the running 
coupling. We employ a second-order Lagrange polynomial in $\ln L$ for 
interpolation of $g_{qq}^{-2}(L)$. The result is shown in
Fig.~\ref{fig:g} as well. The $\beta$ function is then obtained from
\begin{equation}
\frac{1}{2}\frac{\partial\,g_{qq}^{-2}(L)}{\partial\,\ln\,L} =
\bar{\beta}_{qq}(g_{qq}(L)) \,, 
\label{dinv}
\end{equation}
where $\displaystyle \bar{\beta}(g)=g^{-3}\,\beta(g)$. 

\begin{figure}[b!]
\begin{center}
\epsfig{file=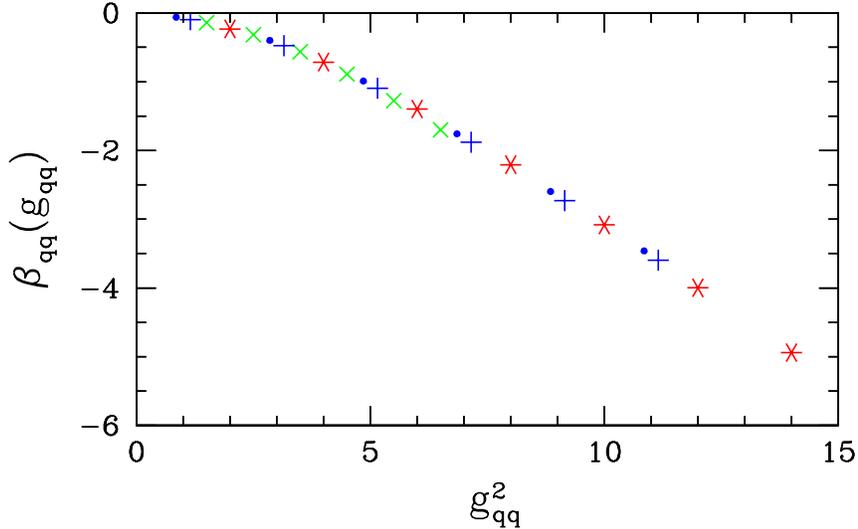,width=12.0cm,clip=true}
\end{center}
\vspace*{-1cm}
\caption{The full $\beta$ function for $N=20$ and $L=\sqrt{6}$
($\times$), $L=\sqrt{9}$ ($+$) and $L=\sqrt{12}$ ($\hexstar$), 
together with the $\beta$ function truncated at $N=15$ for
$L=\sqrt{9}$ ($\bullet)$. The bare coupling has been limited to
$g_0^2\le 1.04$.} 
  \label{fig:c}
\end{figure}

The first two coefficients of the $\beta$ function can directly be
read off from the perturbative expansion of $\displaystyle
\bar{\beta}_{qq}$ in powers of $g_0^2$,
$\displaystyle \bar{\beta}_{qq} = -\left(\beta_0  +\beta_1 g_0^2 +
\cdots\right)$, with $\displaystyle \beta_0=(1/2)\,\partial\, c_2/\partial
\ln L$ and $\displaystyle \beta_1=(1/2)\,\partial\, (c_3-c_2^2)/\partial
\ln L$. The renormalization group predicts that both $c_2$ and
$(c_3-c_2^2)$ are linear functions of $\ln L$. The
first coefficient turns out to be $\beta_0=11.8/(4\pi)^2$, independent
of $L$, as expected. The second coefficient, $\beta_1$, is somewhat
special. It is a
small difference of large numbers, with the condition that the
quadratic terms $\propto \ln^2 L$ in $c_3$ and $c_2^2$ cancel. The 
cancellation is not perfect, which makes $\beta_1$ depend on $L$. At
$L=\sqrt{6}$, the midpoint, we find 
$\beta_1=115/(4\pi)^4$. At this point (\ref{dinv}) coincides with the 
textbook central derivative. Alternatively, we may fit a linear
curve to $(c_3-c_2^2)$. A weighted fit gives
$\beta_1=141(90)/(4\pi)^4$, with a correlation coefficient of
$r=-0.99$, indicating that the (two) fit parameters are strongly
correlated.\footnote{We have not attempted a correlated fit, which we 
  do not consider very meanigful in this case.}  As an estimator for
the weight factor we 
have used the systematic error of $g_{qq}^2(\bar{L})$, which is
estimated to be $\propto 1/\bar{L}^2$ (mod logs, see
(\ref{rc}) {\it et seq.}). The higher coefficients of $\beta_{qq}$ are
no longer linear functions of $\ln L$. 

\begin{figure}[t!]
\begin{center}
\epsfig{file=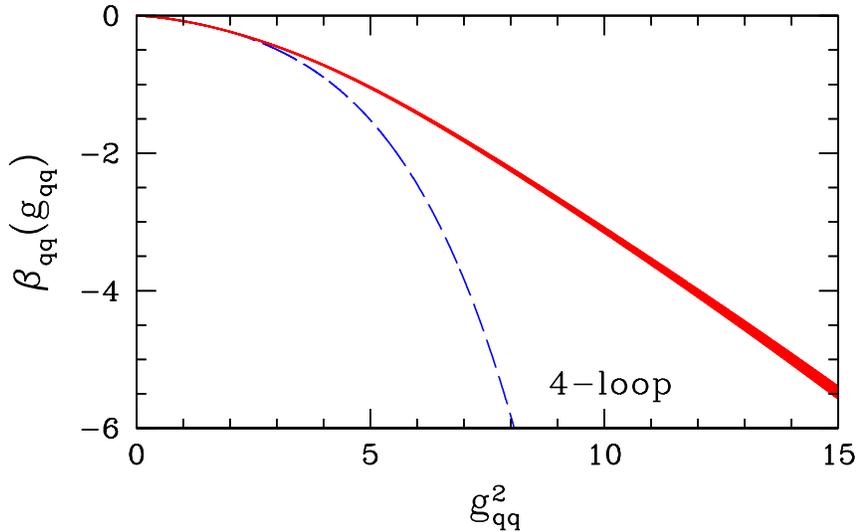,width=12.0cm,clip=}
\end{center}
\vspace*{-1cm}
\caption{The $\beta$ function $\beta_{qq}$ against the coupling
  $g_{qq}^2$. The solid band shows the lattice result, including the
  error. The dashed curve shows the analytic four-loop result.}  
\label{fig:b}
\end{figure}

We now turn to the full $\beta$ function. Sources of error are
discretization effects and malconvergence of the perturbative
series. To test for
possible discretization errors, we compare $\beta_{qq}(g_{qq})$ for
$L=\sqrt{6}$, $\sqrt{9}$ and $\sqrt{12}$ in Fig.~\ref{fig:c}. We do
not see any significant dependence on $L$. To test whether the
perturbative series has 
converged, we compare $\beta_{qq}(g_{qq})$ for $N=20$ and $N=15$. We
see no difference either. We start to see a difference only when the
series is truncated at $N\approx 10$. This indicates that the $\beta$
function is not sensitive to very large ($N\gtrsim 10$) loops, as long
as we keep the bare coupling below $g_0^2 \approx 1.04$.  

\begin{figure}[t!]
\begin{center}
\epsfig{file=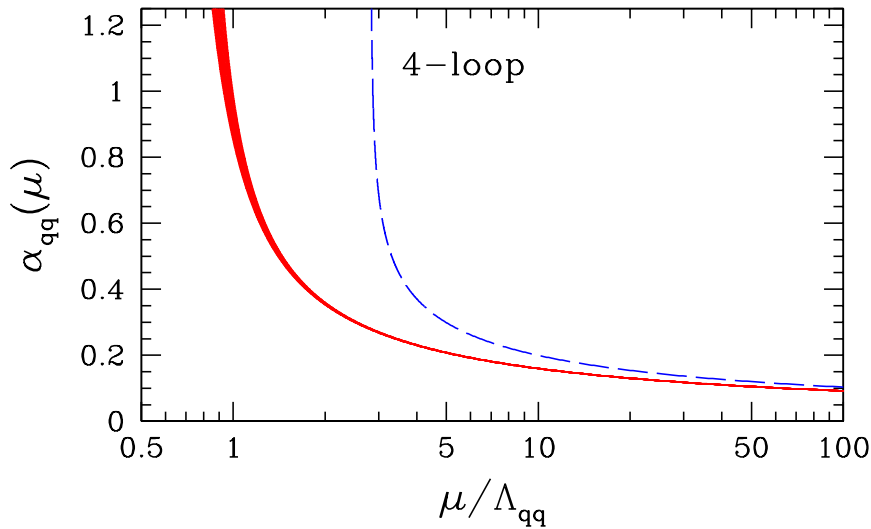,width=12.0cm,clip=}
\end{center}
\vspace*{-1cm}
\caption{The running coupling $\alpha_{qq}(\mu)$ as a function of
  $\mu/\Lambda_{qq}$. The solid band shows the lattice result,
  including the error. The dashed curve shows the analytic four-loop
  result.}    
\label{fig:a}
\end{figure}

In Fig.~\ref{fig:b} we plot our final result for the $\beta$
function. The error band shows the variance of $\beta_{qq}(g_{qq})$ as
$L$ is varied between $L=\sqrt{6}$ and $\sqrt{12}$.\footnote{This is
based on a Pad\'e fit of the form (\ref{pade}) to the lattice data
with $g_0^2 \leq 1.04$.} We compare our result with the analytic
four-loop formula~\cite{Donnellan:2010mx}. The difference grows
rapidly with $g_{qq}^2$. At $g_{qq}^2 = 6.3$ ($\alpha_{qq}=0.5$) the full
$\beta$ function is about half the size of the four-loop analytic result,
and at $g_{qq}^2=8.2$ ($\alpha_{qq}=0.65$) it is one third the size
only. For want of an analytic expression, the  
lattice $\beta$ function can be very well described by the $[3,3]$
Pad\'e approximant 
\begin{equation}
\beta_{qq}(g_{qq})=-g_{qq}^3\, \left(\frac{\beta_0+a_1\, g_{qq}^2 +
  a_2\,g_{qq}^4 +a_3\, g_{qq}^6}{1+b_1\, g_{qq}^2 + b_2\,g_{qq}^4
  +b_3\, g_{qq}^6}\right) \,.
\label{pade}
\end{equation}   
It allows $\beta_{qq}(g_{qq})$ to evolve asymptotically with any power
of $g_{qq}$ from $-3$ to $+9$ and have several zeroes and poles. Using
MINUIT, we fit (\ref{pade}) to the lattice $\beta$ function at
$L=\sqrt{9}$ with $\beta_0\, (=0.074724)$ and $a_1-\beta_0
b_1=\beta_1\, (=0.004612)$ fixed at the one- and two-loop values. The
fit gives $a_2=-0.008910$, $a_3=0.001550$, $b_1=1.3008$, $b_2=-0.1605$,
$b_3=0.0200$. The difference between the lattice result and the fit is
practically invisible. We find that (\ref{pade}) has no poles and no zeroes
on the positive real axis, in contrast to Pad\'e fits to the four-loop
$\beta$ function~\cite{Chishtie:2000eq}. Instead, (\ref{pade}) has one
pole on the negative real axis and two poles deep in the complex. The
same applies to the zeroes of the $\beta$ function. Solving (\ref{mu})
for $g_{qq}^2(\mu)$, we obtain the running coupling 
$\displaystyle \alpha_{qq}(\mu) = g_{qq}^2(\mu)/4\pi$ shown in
Fig.~\ref{fig:a}. The interesting result is that $\alpha_{qq}(\mu)$
hits a wall at $\mu/\Lambda_{qq} \approx 0.7$, indicating that
$\mu$ cannot be taken lower than $\approx 0.7 \Lambda_{qq}$. The
lambda parameter in the force scheme is $\displaystyle \Lambda_{qq} =
1.048 \,\Lambda_{\overline{MS}}$. Thus, $\alpha_{qq}$ and 
$\alpha_{\overline{MS}}$ lie close
together. From~\cite{Gockeler:2005rv} we obtain $\Lambda_{qq}=254(2)\,
\mbox{MeV}$, taking $r_0=0.5\,\mbox{fm}$ to set the scale.  

%
\begin{figure}[h!]
\begin{center}
\epsfig{file=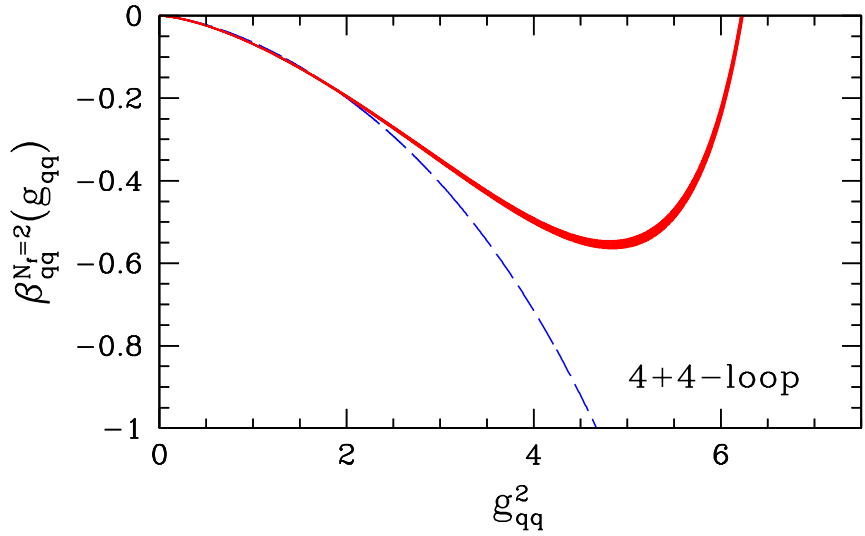,width=12.0cm,clip=}
\end{center}
\vspace*{-1cm}
\caption{The $\beta$ function $\beta_{qq}^{N_f=2}$ against the
  coupling  $g_{qq}^2$. The solid band shows the result of the Pad\'e
  fit, including the error of the pure gauge part. The dashed
  curve shows the analytic $4 + 4$-loop result.}    
\label{fig:b2}
\vspace*{1cm}
\end{figure}
%
\begin{figure}[h!]
\vspace*{0.5cm}
\begin{center}
\epsfig{file=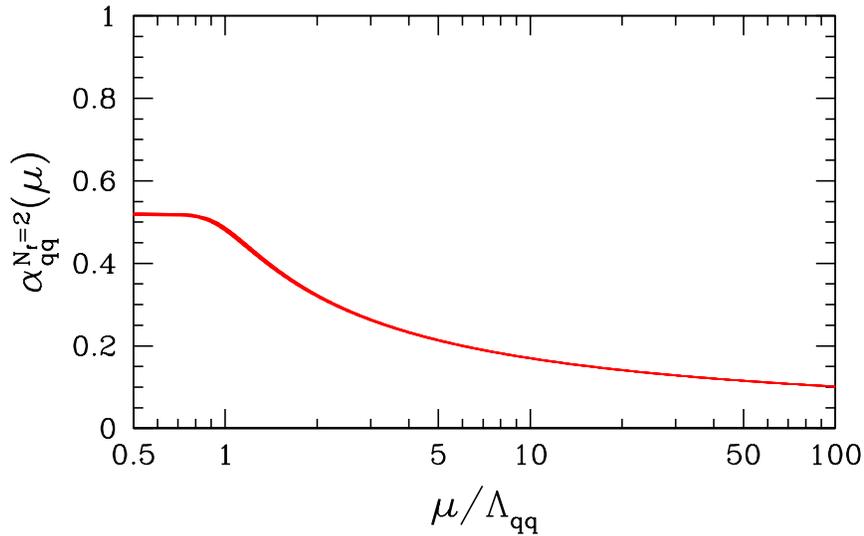,width=12.0cm,clip=}
\end{center}
\vspace*{-1cm}
\caption{The running coupling $\alpha_{qq}^{N_f=2}(\mu)$ as a function of
  $\mu/\Lambda_{qq}$, including the error.} 
\label{fig:a4}
\end{figure}
%

It is tempting now to include fermions. The contribution of massless
fermions is known to four loops~\cite{Donnellan:2010mx}. Adding
together the gluonic and fermionic contribution, we arrive at the
$\beta$ function for $N_f$ quark flavors
\begin{equation}
\begin{split}
\beta_{qq}^{N_f}(g_{qq}) = \beta_{qq}(g_{qq}) &- g_{qq}^3
\Big[\beta_0^{N_f} + \beta_1^{N_f}\,g_{qq}^2 +
\beta_2^{\,qq,N_f}\,g_{qq}^4 \\[0.25em]
&+ \beta_3^{\,qq,N_f}\,g_{qq}^6 +
\beta_{3,l}^{\,qq,N_f}\,g_{qq}^6\,\ln{\left(3g_{qq}^2/8\pi\right)}\Big] \,, 
\end{split}
\label{pf}
\end{equation}
where $\beta_i^{\,(qq,)N_f}$, $i=0, 1, 2$ and $3$, are the one-, two-,
three- and four-loop coefficients, respectively, of the
fermionic part of the $\beta$ function in the $qq$ scheme, and
$\beta_{3,l}^{\,qq,N_f}$ is the coefficient of the four-loop
logarithmic contribution.
We are interested in the low-energy behavior of
$\alpha_{qq}(\mu)$. This is governed by the $u$ and $d$ quarks, which
can be assumed to be massless. We thus are led to consider the case
$N_f=2$. In view of 
successful predictions of higher-order contributions in the
past~\cite{Ellis:1997sb}, we fit a $[3,3]$ Pad\'e approximant
to (\ref{pf}). The result of the fit is
\begin{equation}
\begin{split}
\beta_{qq}^{N_f=2}(g_{qq}) = &-g_{qq}^3\,\Big(\frac{0.066281 +
  0.090111\, g_{qq}^2  - 0.010112\, g_{qq}^4 - 0.000857\, g_{qq}^6} {1
  + 1.3053\, g_{qq}^2 - 0.1711\, g_{qq}^4 - 0.0041\, g_{qq}^6}\\[0.25em]
  &\hspace*{6.8cm}+ 0.000044\, g_{qq}^6\,\ln\,g_{qq}^2 \Big) \,,
\end{split}
\label{bf2}
\end{equation}
where we have kept the logarithmic contribution separate. This is
justified, as the latter contributes only a few percent in the region
that is of interest to us. The $\beta$
function (\ref{bf2}) is shown in Fig.~\ref{fig:b2}. It has a zero at
$g_{qq}^2 = 6.3$ followed by a pole at $g_{qq}^2 = 7.3$. The
other two poles lie on the negative real axis. The coefficient $a_1$
(in the notation of (\ref{pade})) has changed by $13\%$ (from
$0.101813$ to $0.090111$), while $b_1$ has practically not changed at
all, and the subleading negative coefficients $a_2$ and $b_2$, being
an order of magnitude smaller, have changed by $15\%$ or less, as
compared to (\ref{pade}). For the $[3,3]$ Pad\'e approximant to be
sufficiently well constrained, it was important to know the fermionic
contribution to four loops. From the $\beta$ function (\ref{bf2}), and
(\ref{mu}), we may now compute the running coupling
$\alpha_{qq}^{N_f=2}(\mu)$. The result is shown in
Fig.~\ref{fig:a4}. As expected, the running coupling freezes at
$\alpha_{qq}^{N_f=2} \approx 0.5$ as $\mu$ is taken to zero,
rendering the theory scale invariant.

\begin{figure}[t!]
\begin{center}
\epsfig{file=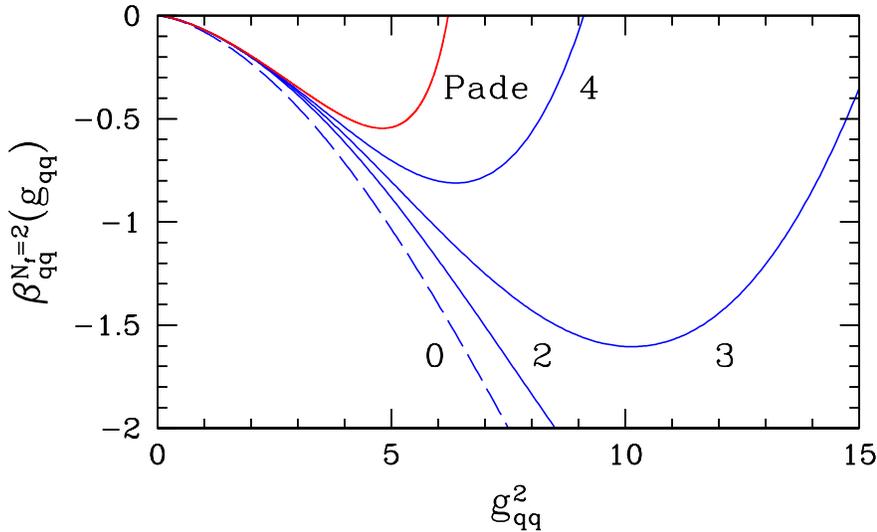,width=12.0cm,clip=}
\end{center}
\vspace*{-1cm}
\caption{The pure gauge $\beta$ function (\ref{pade}) plus the
  fermionic contribution to $0$, $2$, $3$ and $4$ loops, together with
  the Pad\'e approximant of Fig.~\ref{fig:b2}.} 
\label{fig:nf}
\end{figure}

The crucial point is that at larger couplings the full, pure
gauge $\beta$ function is significantly smaller (in absolute terms)
than its four-loop counterpart. That gives the fermionic part
considerably more weight. In Fig.~\ref{fig:nf} we show the sum
(\ref{pf}) of gluonic and fermionic contribution, in dependence on the
number of loops of the fermionic part. Already at three loops 
the $\beta$ function shows a second zero, which moves to
$\alpha_{qq}^{N_f=2} \approx 0.7$ at four loops and down to
$\alpha_{qq}^{N_f=2} \approx 0.5$ in case of the Pad\'e approximant
(\ref{bf2}). This votes for the existence of an
infrared fixed point for two massless quark flavors. The exact
position of the second zero is subject to change though. 

The question that arises now is how significant are nonperturbative
contributions. In~\cite{Horsley:2012ra} we have examined the
difference $\Delta W(L,T)$ of nonperturbative (Monte Carlo) and
perturbative Wilson loops $W(L,T)$ of size $L,T \leq 2$. We find that
the contributions  
not accounted for by the perturbative series are way smaller than
$\sum_{n=16}^{20} r_n(L,T)\, g_0^{2n}$, the difference between the
full ($N=20$) and truncated ($N=15$) series, for $0.95 \leq g_0^2 \leq 
1.04$, and thus have no effect on our results (see
Fig.~\ref{fig:c}). 
Below $g_0^2 = 0.95$ we do not know $\Delta W(L,T)$,
but it is expected that it drops to zero with some power of the lattice
constant, faster than the perturbative series, as $g_0^2 
\rightarrow 0$. Regarding larger loops, it has been
argued~\cite{Shifman:1980ui} that the 
nonperturbative contribution, or pieces of it, might increase as the
size of the loop is increased. We do not see such a
behavior~\cite{Horsley:2012ra} (Fig.~19, right panel), but cannot
exclude it presently for 
$L, T > 2$. If at all, the argument might apply to quadratic loops,
but certainly not to $T \gg L$. To probe the $\beta$
function in the vicinity of the infrared fixed
point, it will be sufficient to consider Wilson loops of spatial
extent $\approx 0.3\,\mbox{fm}$ (see below).      

To conclude, we have computed the SU(3) pure gauge $\beta$
function from Wilson loops to $20th$ order numerical stochastic
perturbation theory. First results from the $12^4$ lattice are
intriguing. To put our calculations into perspective, at
$g_{qq}^2(L)=6$, corresponding to $L=0.32\,\mbox{fm}$ using
$r_0=0.5\,\mbox{fm}$ to set the scale~\cite{Necco:2001xg}, the lattice
constant varies between $a=0.146\,\mbox{fm}$ and 
$0.093\,\mbox{fm}$. We hope to extend the calculations to  
larger lattices and larger Wilson loops in due
course. This will allow us to take the limit $T \rightarrow \infty$,
and to extend the calculations to smaller (continuum) values of the
bare coupling $g_0^2$. Our 
calculations so far suggest that the $\beta$ function is of
perturbative origin. It needs to be seen if this behavior continues to
hold in the continuum limit. To check that, we intend to compute $\Delta
W(L,T)$ for a few representative loops. 
Last not least, larger lattices will allow us to 
probe the $\beta$ function at even larger values of the coupling. To
corroborate our results on the infrared  
fixed point of the QCD $\beta$ function for a small
number of massless quarks beyond any doubts, we would need to compute
the fermionic contribution to higher loops. That appears to be
feasible. In~\cite{Di Renzo:2004ge} numerical stochastic perturbation
theory has been extended to full QCD, and first results on Wilson loops
have been reported. Perhaps, this is the only possibility of computing
the $\beta$ function for massless quarks at small virtualities.

\section*{Acknowledgement}

This work has been supported in part by the EU under contract 283286
(HadronPhysics3) and DFG under contract SCHI 422/9-1, which we gratefully
acknowledge.

\end{document}